# Amplitude-modulated pulses based phase-sensitive OTDR in distributed sensing

**Qian He*, Heng Jiang, Chengdan Tan, Rong Liu and Lijun Tang**

Changsha University of Science & Technology, Changsha 410114, China;
* Correspondence: elfxixi1021@126.com;

**Abstract:** The detected frequency response range of phase-sensitive optical time domain reflectometry is limited by the sensing fiber length. In this paper, we demonstrate an amplitude modulated pulses based phase-sensitive OTDR to detect vibrations with high spatial resolution and wide frequency response range. The amplitude modulated pulses consisting of narrow pulses and background light are injected into the sensing fiber as probe light. The Rayleigh backscattering generated by the background light interferes at the detection end, presenting a phase sensitive response to external vibrations. The Rayleigh backscattering of narrow pulses are tracked to locate vibrations. The frequency information is obtained from the frequency domain analysis of the interference signal. The system can realize high frequency response with low repetition rate of probe pulses by one-end detection.

---

## 1. Introduction

Fiber optical distributed sensing owns the merits of continuous monitoring, long-haul sensing and immunity to electromagnetic. During the past years, distributed vibration sensing has been studied for health monitoring of large-scale structures and intrusion detection. Vibrations are always related to material damages and man-made activities, such as leakage of high-pressure water/gas pipeline, crack of materials, and intrusion behaviors. Both the accurate location and real frequency response of external vibrations are demanded for fast fault analysis in industries.

Techniques for distributed vibration sensing based on backscattering in sensing fiber are intensively studied because of real-distributed measurements and accurate locations. A system based on Brillouin optical correlation domain analysis (BOCDA) achieved 8.8 Hz frequency response [1]. Later, Brillouin optical time-domain analysis (BOTDA) based configurations for vibration detection are demonstrated, 98 Hz frequency response is fulfilled [2]. A system using sweep-free BOTDA is demonstrated and the frequencies up to 400 Hz are spectrally analyzed [3].

Phase-sensitive OTDR (φ-OTDR) for distributed intrusion detections is proposed. The system is cost-effective and real-time compared to Brillouin scattering based configurations [4]. The frequency response of material cracks ranges from a few kilo Hertz to Mega Hertz. However, the detected frequency response range is limited by the sensing fiber length and data averaging. A coherent φ-OTDR is demonstrated to improve signal-to-noise ratio. The highest frequency response of 1 kHz is realized by using moving averaging and moving differential method [5]. Systems enhance sampling rate by the frequency multiplexing technique are proposed in [6,7]. A large number of carrier frequencies is demanded to further increase the highest frequency response, resulting in a complicatedly de-multiplexing of data analysis. Configurations based on combination of φ-OTDR and MZI are demonstrated to realize up to Mega Hertz frequency response and high spatial resolution [8,9]. While the two-end detection limits real applications.

In this paper, we propose a phase-sensitive OTDR system based on amplitude-modulated pulses to realize wide frequency response range with low repetition rate of probe pulses. The system is simple and cost-effective.

## 2. Principle and Experimental setup

The amplitude-modulated optical pulses are shown in Fig.1. One measurement interval includes an ultra-narrow pulse with pulse-width of t1 and a quasi continuously background light with lasting time of t2. The Rayleigh backscattering signal IP (t, z) of the narrow pulses are detected to obtain vibration positions. The Rayleigh backscattering IB (t, z) of the background light is phase modulated by the external vibration and interferes at the detection end of the sensing fiber.

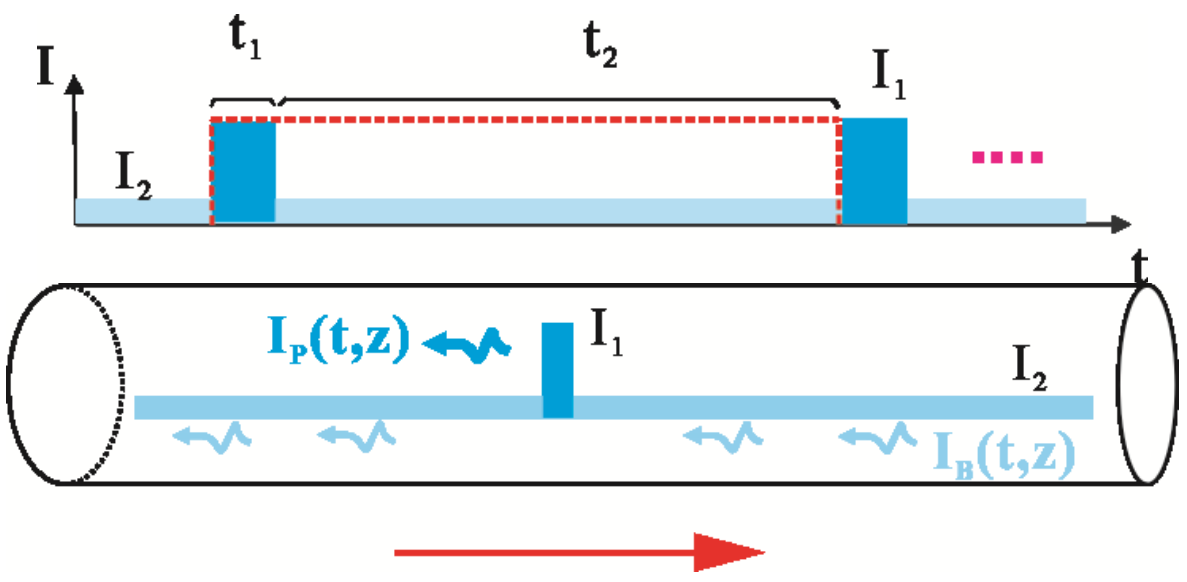

Fig.1 The amplitude-modulated pulses.

The backscattering signal IB (t, z) induced by the background light is used to extract frequency response of external vibrations. Noticed that the received backscattering signals IB interfere at the detection end, the sensitivity of the sensing system is further enhanced.

Figure 2 shows the experimental setup. A laser working at 1550.01 nm with less than 100 kHz line-width and 10 mw output power (ID Photonics CoBriteH01) is used as the coherent light source. The CW laser is intensity modulated by an acoustic optic modulator (AOM) driven by a signal generator. The modulated optical pulses are then amplified by an erbium doped fiber amplifier (EDFA, Amonics AEDFA-23-B) and filtered by a fiber Bragg grating (FBG) to eliminate the spontaneous emission noises. After that, the modulated pulses are injected into the sensing fiber (Corning SM-28e) through the optical circulator 2.

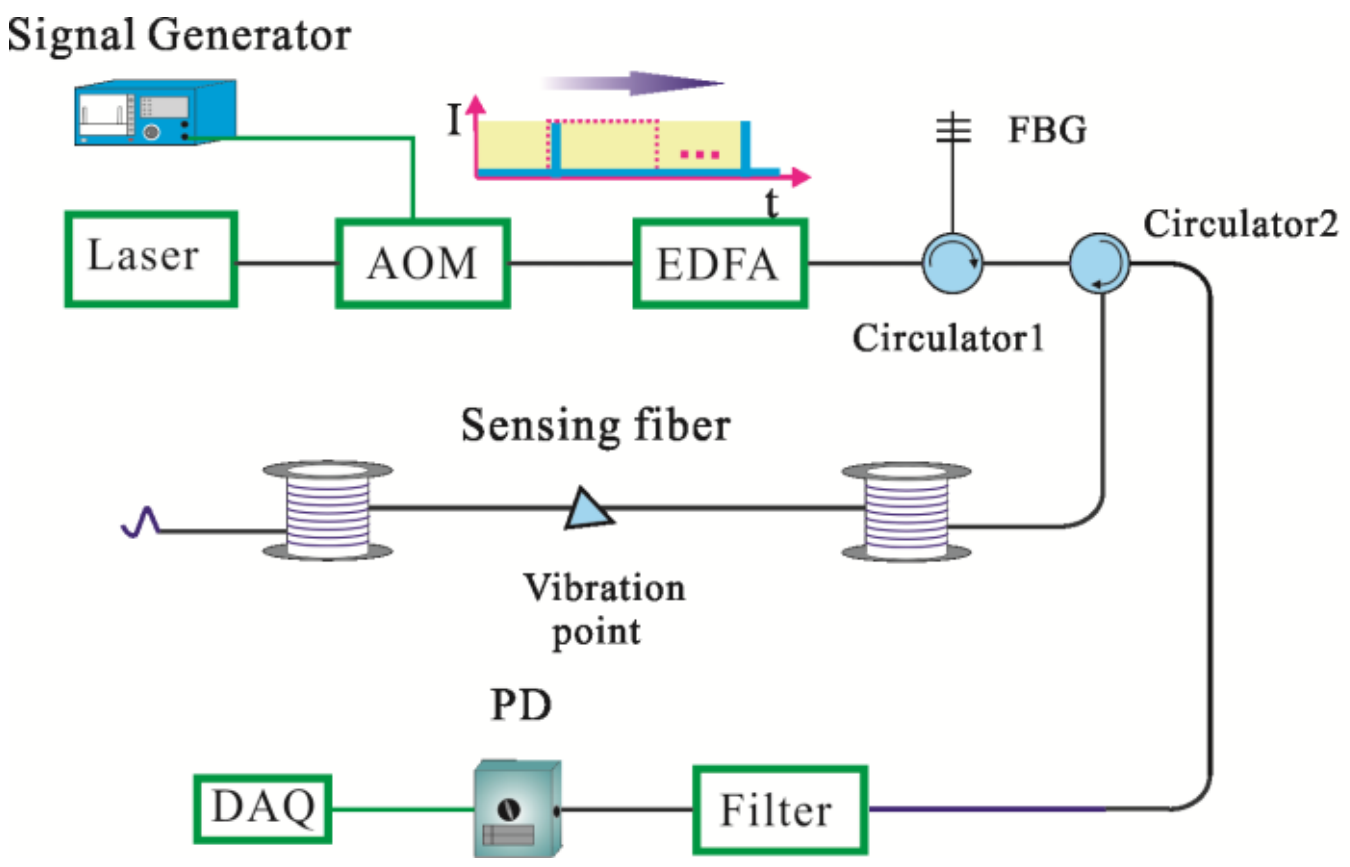

Fig.2 Experimental setup (green lines: electrical paths; dark lines: optical paths).

The Rayleigh backscattering generated by the amplitude-modulated pulses travels back and are redirected into one port of a photodetector (PD, Thorlabs PDB450C) by the circulator. The electrical signals are then sampled by a data acquisition card (DAQ card) with 100 MS/s sampling rate.

## 3. Experimental results and Discussions

The AOM is driven by electrical modulated-pluses with 1V of high voltage and 0.2V of background voltage.  Amplitude-modulated pulses with repetition rate of 5 kHz and pulse-width of 50 ns are injected into the sensing fiber as probe pulses.

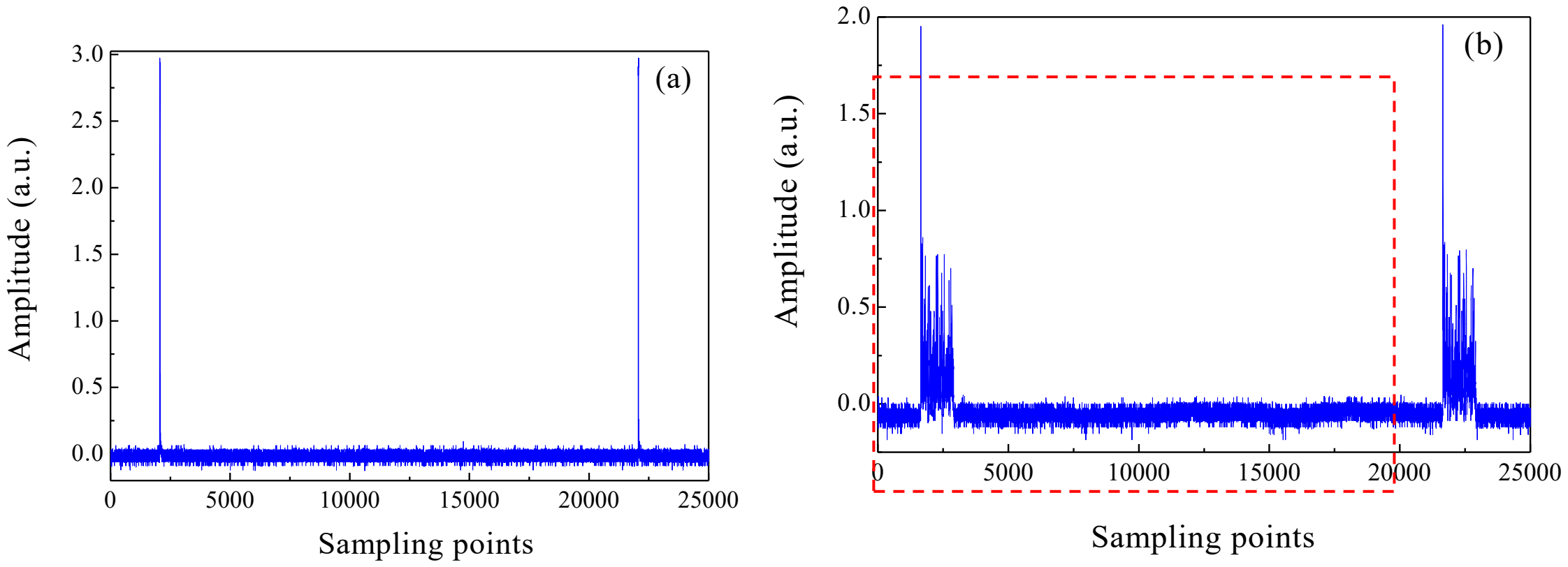

Fig.3 (a) Transmitted probe light at sensing fiber end. (b) The Rayleigh backscattering signals of the probe pulses.

The ~20000 sampling points between the two measurement intervals correspond to the repetition time of 0.2 ms. Figure 3 (a) shows the transmitted probe light at the sensing fiber end. Figure 3 (b) gives the sampled backscattering trace at the detection end. One measurement interval is marked with the red dashed line.

A piezo-electric transducer tube (PZT) is arranged at ~925 m of the sensing fiber with 0.18 m of sensing fiber wounded on. After sampling 100 consecutive intervals, differentials between two adjacent backscattering traces are carried out to locate the vibration position. Figure 4 shows the vibration location along the sensing fiber.

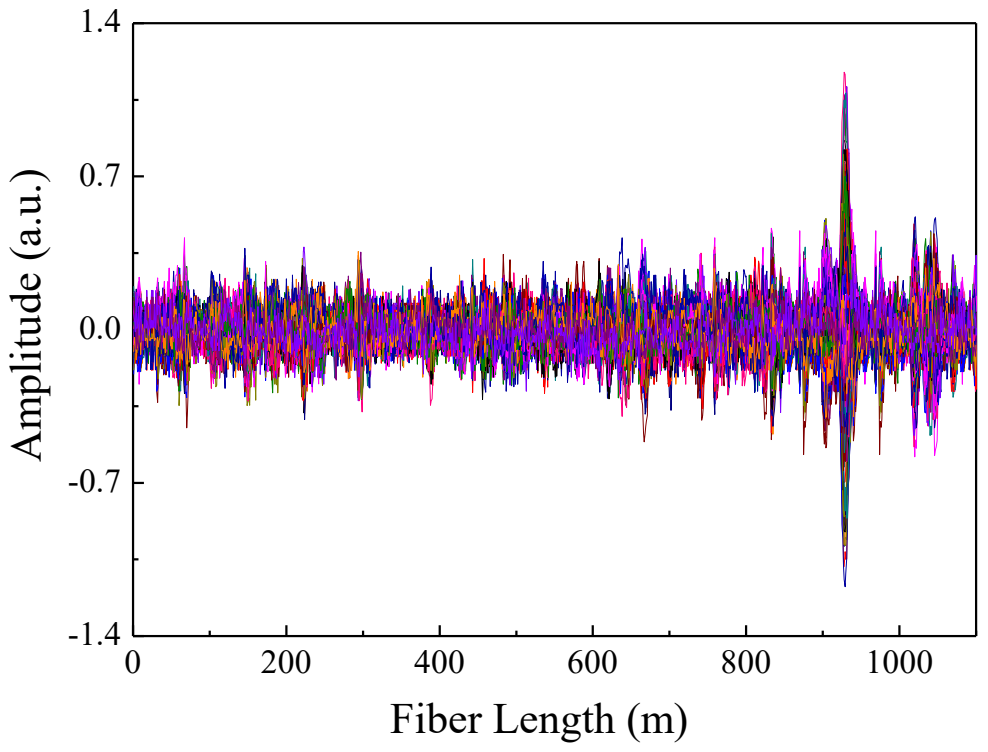

Fig.4 Superposed differential traces show the vibration position at ~925 m.

The measurements of continuous vibrations with frequency of 21 kHz are carried out. Figure 5 (a) gives the backscattering signal with modulated amplitude. The modulation in background light can be observed in the red dashed line. The sampled traces are processed by Labview software. 100 consecutive backscattering traces are used for FFT analysis, and the power spectrum is given in Fig. 5 (b). The peak appears at 21 kHz indicating the detected vibration frequency. The harmonic components of 5 kHz are induced by the repetition rate of probe pulses. In order to eliminate the harmonic components, the backscattering signals generated by narrow pulses are removed from the original sampling traces. The frequency response of vibrations with 21 kHz after removing the harmonic components is shown in Fig. 6.

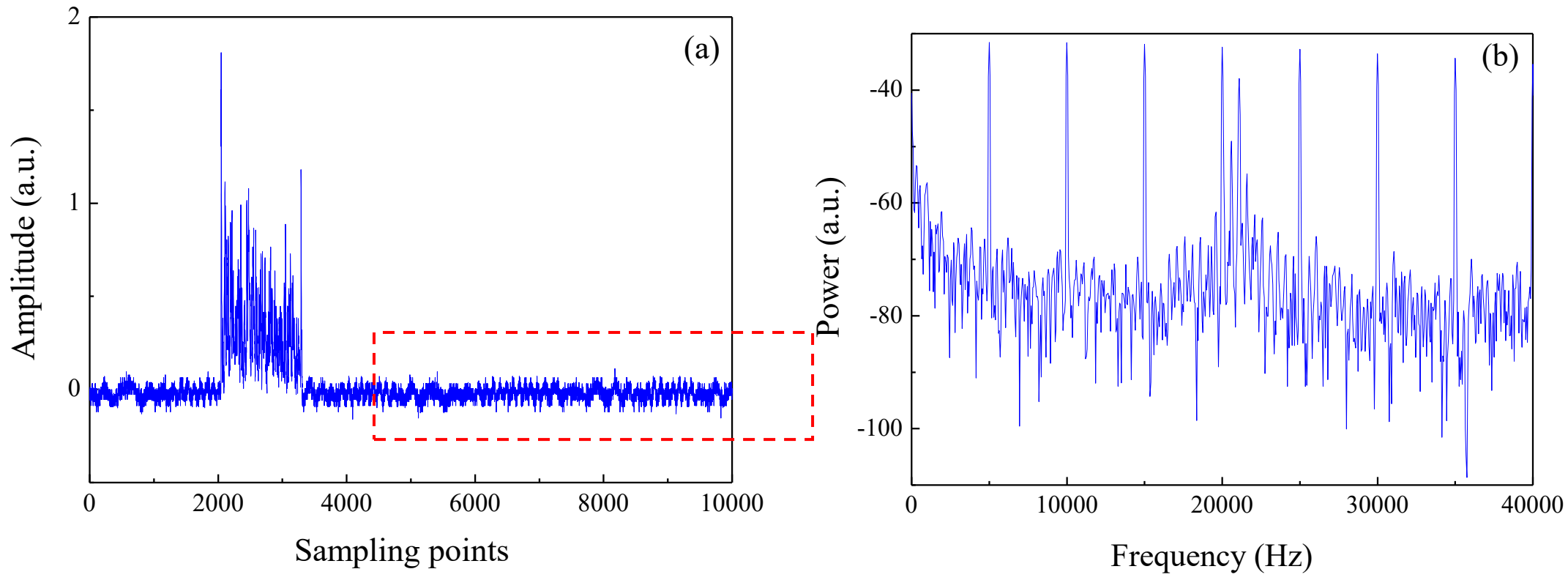

Fig.5 Experimental results of vibrations with frequency of 21 kHz: (a) the Rayleigh backscattering traces; (b) the frequency response of vibrations.

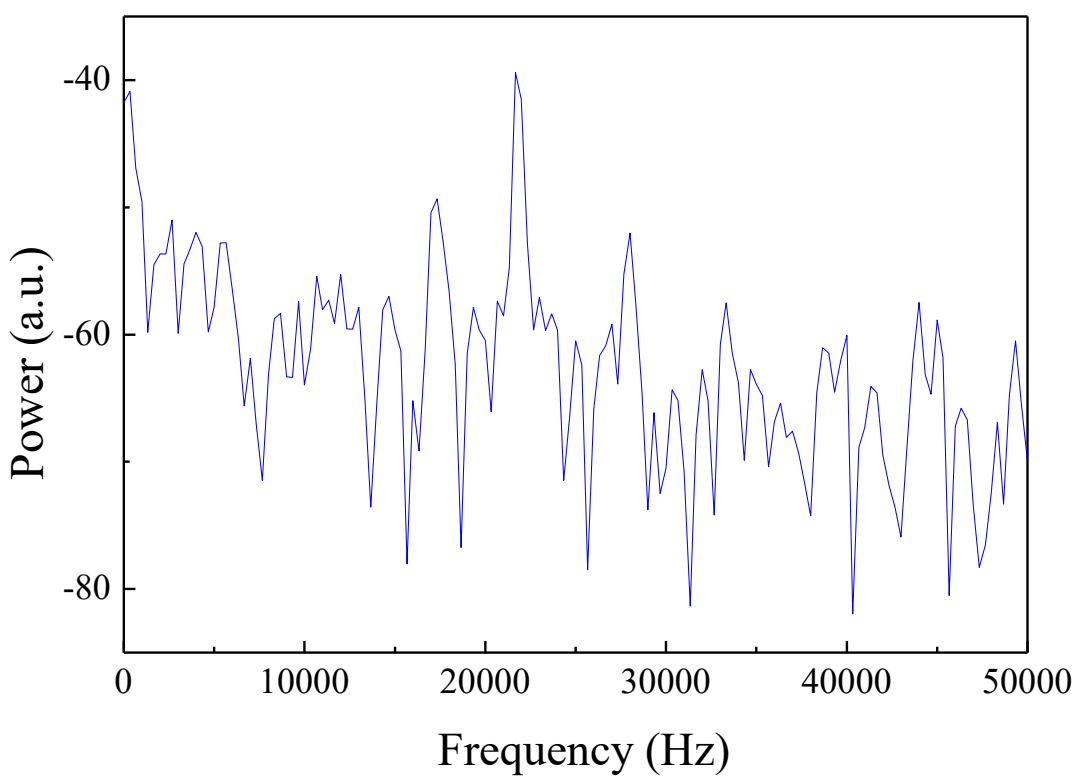

Fig.6 Frequency response of vibrations with frequency of 21 kHz after removing backscattering signals generated by narrow pulses.

In real applications, the occurrence and end of vibrations are unpredictable. Measurements of the PZT working in burst mode are carried out. The PZT is driven by a burst sinusoidal signal with frequency of 22 kHz and lasting time of 10 ms. The sampled backscattering signal is given in Fig. 7 (a). The burst vibration started from sampling points of ~65200. After removing the backscattering signal induced by narrow pulses from the original sampling traces, the frequency response of the burst vibration is analyzed and given in Fig. 7 (b).

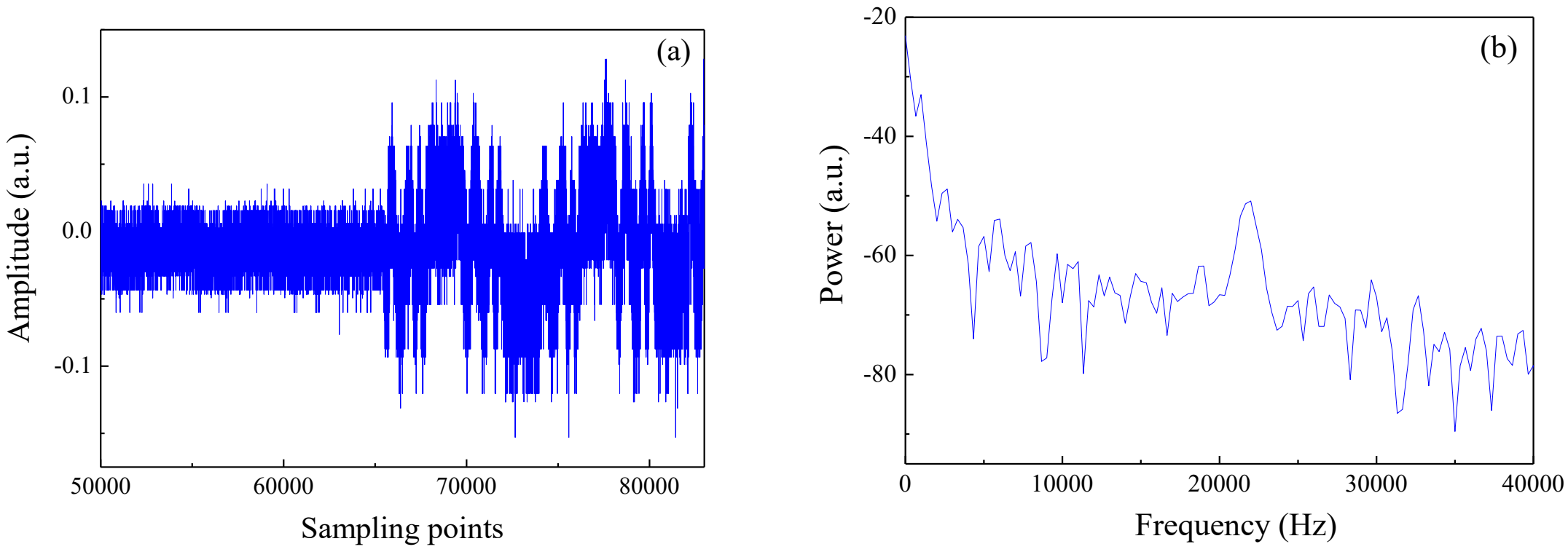

Fig.7 (a) Backscattering signal shows the process of vibration. (b) Frequency response of the burst vibration.

In order to realize a single-end sensing system, the backscattering signals generated by narrow pulses and background light are both travelling back and received by the photoelectric detector at the detection end. The increase of the background light power directly enhances the SNR of frequency response. However, the increased backscattering intensity of background light results in lower SNR of the position signal. We define IR as the intensity ratio of the narrow pulses and the background light. The SNR of vibration location versus IR is studied and illustrated in Fig.8. Experimental results show that there is a trade-off between the SNR of location and the SNR of frequency response. The intensity ratio IR can be set according to actual demands.

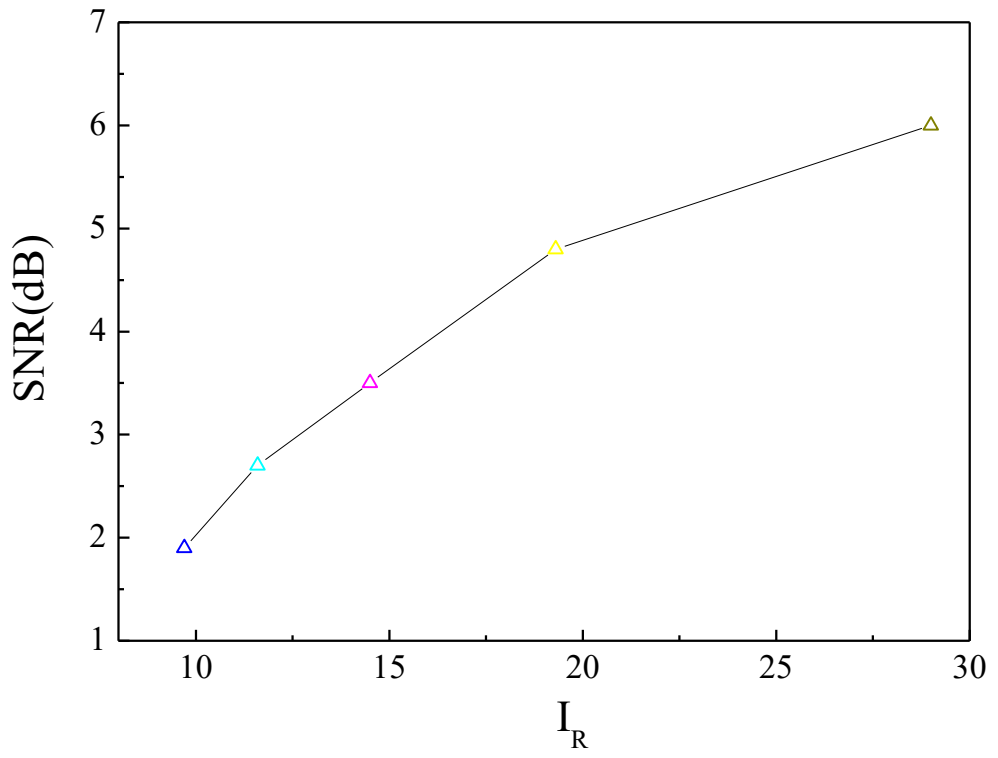

Fig.8 The relationship of IR and the SNR of position signals.

The repetition rate of probe pulses is expected to reach the limit in conventional φ-OTDR to acquire high frequency response. In the proposed system, since the frequency response is extracted from the interference of Rayleigh backscattering generated by the background light, the repetition rate of the modulated pulses is not critically required. Therefore, the device cost is reduced effectively. Moreover, the one-end detection system is practical in applications.

## 4. Conclusions

In this paper, we demonstrate an amplitude-modulated-pulse based φ-OTDR system to detect vibrations with high spatial resolution and wide frequency response range. Amplitude modulated pulses consisting of narrow pulses and background light are injected into the sensing fiber as probe light. The Rayleigh backscattering of narrow pulses are tracked to locate vibration points. The Rayleigh backscattering induced by background light carries the frequency response and interferes at the detection end. Experimental results show that frequency response of 22 kHz with 5 kHz repetition rate of modulated pulses is detected. The system can realize high frequency response, one-end detection and cost-effective in distributed vibration sensing.

## 5. Conclusions

This section is not mandatory, but can be added to the manuscript if the discussion is unusually long or complex.

**Author Contributions:** conceptualization, Qian He.; methodology, Heng Jiang and Chengdan Tan.; formal analysis, Rong Liu; writing—original draft preparation, Qian He and Heng Jiang.; writing—review and editing, Lijun Tang.;

**Conflicts of Interest:** The authors declare no conflict of interest.